\documentstyle[aps,multicol,epsf]{revtex}

\newcommand{\beq}{\begin{equation}}
\newcommand{\eeq}{\end{equation}}
\newcommand{\beqa}{\begin{eqnarray}}
\newcommand{\eeqa}{\end{eqnarray}}
\newcommand{\ket}[1]{| #1 \rangle}
\newcommand{\bra}[1]{\langle #1 |}

\newcommand{\opa}{\hat{A}}
\newcommand{\opb}{\hat{B}}

\begin{document}

\draft
\title{Complementarity and the uncertainty relations}

\author{Gunnar Bj\" ork,\thanks{Electronic address:
gunnarb@ele.kth.se} Jonas S\"
oderholm, Alexei Trifonov,\thanks{Permanent address: Ioffe
Physical Technical Institute, 26
Polytekhnicheskaya, 194021 St. Petersburg, Russia.} Tedros Tsegaye,
and Anders Karlsson}

\address {Department of Electronics, Royal Institute of
Technology (KTH), Electrum 229, SE-164 40 Kista, Sweden}

\date{\today}
\maketitle

\begin{abstract}
We formulate a general complementarity relation
starting from any Hermitian operator with discrete
non-degenerate eigenvalues. We then elucidate the
relationship between quantum complementarity
and the Heisenberg-Robertson's uncertainty relation.
We show that they are intimately connected. Finally we
exemplify the general theory with some specific
suggested experiments.
\end{abstract}



\begin{multicols}{2}

\section{Introduction}

A fundamental notion of quantum mechanics is
complementarity which expresses the fact that any
quantum system has at least two properties that cannot
simultaneously be known. One of these
complementary property pairs, and perhaps the
historically most important, is the wave-particle
duality. A quantum system has both particle-like and
wave-like properties. However, observation of one
property precludes the observation of the other.
Recently several quantitative expressions of this
specific duality were derived
\cite{Wootters,Greenberger,Mandel,Walther,Raymer,Kwiat1,Tan,Kwiat2,Jaeger,Englert,Bjork},
Some of these expressions have been experimentally
confirmed \cite{Herzog,Rempe}. In this paper we will
follow Englert's \cite{Englert} definitions most
closely. In general, inequalities quantifying the
wave-particle duality are posed in the context of
interferometry. In any single particle interferometer
it is meaningful both to ask which of the
interferometer paths the particle took, and to record
the visibility of a large number of identically
prepared systems. In this paper we shall discuss the
notions of which path and visibility in a general
framework as to encompass {\em any} system defined in
a two-dimensional Hilbert-space.

Let us begin by establishing some notation. Assume that
we would like to estimate which of two paths, call them
$+$ and $-$, a particle took. The only information we
have are known probabilities $w_+$ and $w_- = 1-w_+$
for the two events. The Maximum Likelihood
(ML) estimation strategy (which is one of many possible
strategies) dictates that we should, for each and every
event, guess that the particle took the most likely
path. The strategy will maximize the likelihood $L$ of
guessing correctly. The likelihood will be $L = {\rm
Max} \{ w_+, w_-\}$, and from this relation it is
evident that $1/2 \leq L \leq 1$. The likelihood can be
renormalized to yield the predictability $P$
\cite{Englert}, given by 
\beq
P = 2 L -1 .
\label{eq:disting}
\eeq 
It is clear that $0 \leq P \leq 1$,
where $P=0$ corresponds to a random guess of which path
the particle took, and $P=1$ corresponds to absolute
certainty about the path. Take note that
$P$ corresponds to the likelihood of the correct
{\em estimated outcome}. If one were to estimate the
path of an ensemble of identically prepared particles
according to the ML strategy one should guess
identically the same path for each system. That is,
every estimate would be identical so that the
estimated path would have no variance. If, on the
other hand, one made a {\em factual measurement} of
the path, one would get a random outcome characterized
only by the probabilities
$w_+$ and $w_-$.

One could also measure the visibility when the two
path probability amplitudes interfere. The visibility
$V$, too, is a statistical measure which requires an
ensemble of identically prepared systems to
estimate. The classical definition of $V$ is 
\beq V = {I_{max}-I_{min} \over I_{max}+I_{min}} ,
\label{eq:classical visibility definition}
\eeq where
$I_{max}$ and $I_{min}$ are the intensities of the
interference fringe maxima and minima. For a single
particle we can only talk about the probability $p$ of
the particle falling on a specific location on a
screen, or exiting one of two interferometer ports.
(Do not confuse this probability $p$ with the
predictability $P$.) The probability
$p$ will vary essentially sinusoidally with position
on the screen, or with the interferometer arm-length
difference. In this case the natural definition of
$V$ is
\beq V = {p_{max}-p_{min} \over p_{max}+p_{min}} ,
\label{eq:quantum visibility definition}
\eeq

It has been shown \cite{Englert} that $P$ and $V$ for a
single particle, satisfies the following inequality:
\beq
P^2 + V^2 \leq 1 ,
\label{eq:Complementarity relation}
\eeq
where the upper bound is saturated for any pure state.
A relevant question to ask is how this inequality is
related to the uncertainty principle, which is also  a
quantitative inequality manifesting complementarity.
Furthermore one can ask what observables, if any,
$P$ and $V$ correspond to? We shall attempt to
clarify these issues in this paper. We shall also
show that a relation of the same form as
(\ref{eq:Complementarity relation}) can be derived for
any Hermitian operator by constructing a complementary
(and therefore non-commuting) Hermitian operator.
Finally we shall see that there is a substantial
difference between the non-simultaneous and the
simultaneous Heisenberg-Robertson uncertainty
relations.

\section{Generalized complementarity}

Let us consider the states $\ket{A_+}$ and $\ket{A_-}$,
which are eigenstates of the Hermitian operator
$\hat{A}$ with eigenvalues $A_+ \neq A_-$,
respectively. Therefore, $\langle A_+ | A_- \rangle =
0$. In the following we shall assume that the
two-dimensional Hilbert-space spanned by these two
orthonormal states is sufficient to describe the
system. It is permissible that the operator $\opa$ has
additional eigenstates, but to analyze the
complementarity relation we shall only consider
transitions between, and hence superpositions of, two
of the eigenstates. Below, for the sake of clarity, we
shall refer to the {\em system mode}, which is the
physical entity one can make measurements on, and the
{\em system state}, which is the quantum-mechanical
state of the mode, i.e. a result of a measurement. It
is rather straightforward to extend the relation
(\ref{eq:Complementarity relation}) to the case where
a larger Hilbert-space is needed
\cite{Bjork}, but we will refrain from attempting such
a generalization since two states are sufficient to
elucidate the answers to the questions posed above. We
shall furthermore assume that our system is prepared in
a general state with the associated density operator
\beq
\hat{\rho} = \left [
\matrix{
w_+ & \rho_{12} e^{-i \theta} \cr
\rho_{12} e^{i \theta} & w_-}
\right ],
\label{eq:state definition}
\eeq
where $w_+$, $w_-=1-w_+$, $\rho_{12} \leq \sqrt{w_+
w_-}$ and $\theta$ can be assumed to be real positive
numbers without any loss of generality, and where the
density operator is expressed in matrix form in the
$\ket{A_+}$ and
$\ket{A_-}$ basis. We can identify the parameter $w_+$
as the {\em a priori} probability of finding the
system in the state $\ket{A_+}$, and similarly for
$w_-$.

If we use the maximum likelihood strategy to predict
the outcome of a measurement of $\hat{A}$, then we will
succeed with the likelihood $L={\rm Max}\{w_+,w_-\} =
(w_+ + w_- + |w_+ - w_-|)/2 = (1+|w_+ - w_-|)/2$, where
${\rm Max}\{w_+,w_-\}$ denotes the maximum of the two
probabilities $w_+$ and $w_-$. From this equation, and
(\ref{eq:disting}) follows that
\beq
P = |w_+ - w_-| = \sqrt{(1-2w_-)^2}= \sqrt{1-4w_+ w_-}
.
\label{eq:predictability}
\eeq
Note that this quantity is based only on the
probabilities
$w_+$ and $w_-$ which characterize the {\em
preparation} of the state $\hat{\rho}$. 

\begin{figure}
\leavevmode
\epsfxsize=8cm
\epsfbox{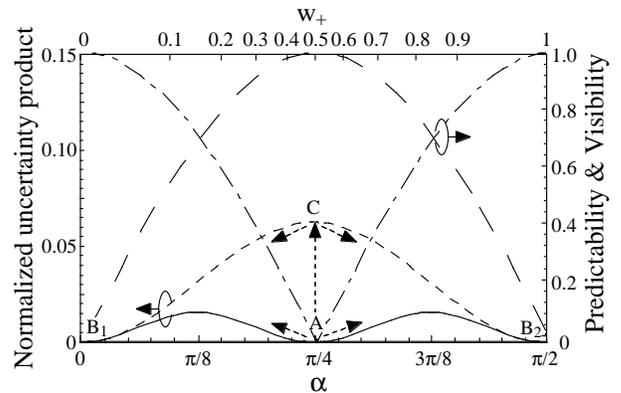}
\vspace{0.3cm}
\caption{The normalized dimensionless minimum
uncertainty product of the operators $\opa$ and
$\opb$ as a function of $w_+=\sin^2(\alpha)$. The
dashed line represents the uncertainty product of the
second class of intelligent states. The uncertainty
product of the initial state for operators
$\opa$ and $\opb$ is bounded from below by the solid
line and from above by the dashed line. The dot-dashed
line represents the predictability of the state and
the long dashed line the visibility of a pure state.}
\label{fig:Uncertainty product}
\end{figure}

Next consider the two unitary operators
\beq
\hat{U}_{PS} = \left [ 
\matrix{
1 & 0 \cr
0 & \exp(i \phi)}
\right ] ,
\label{eq:ups}
\eeq
and
\beq
\hat{U}_{BS} = \left [ 
\matrix{
\cos(\xi) & i \sin(\xi) \cr
i \sin(\xi) & \cos(\xi)}
\right ] ,
\label{eq:ubs}
\eeq
which both are expressed in matrix form in the
$\ket{A_+}$ and $\ket{A_-}$ basis. The two unitary
transformations correspond to a generalized phase-shift
(of state $\ket{A_-}$) and a generalized
``beamsplitter'', respectively. The density matrix of
the unitarily transformed state is $\hat{\rho}' =
\hat{U}_{BS} \hat{U}_{PS}
\hat{\rho} \hat{U}_{PS}^\dagger \hat{U}_{BS}^\dagger$.
The probability of obtaining the outcome $A_+$ from
$\hat{\rho}'$ if
$\hat{A}$ is measured is $\bra{A_+} \hat{\rho}' \ket{A_+}$.
We denote the maximum and the minimum of this
probability, as a function of $\phi$ and $\xi$,
$p_{max}$ and $p_{min}$. One finds that the global
maxima and minima are found for $\xi = \pi/4 + n
\pi/2$, where $n$ is an arbitrary integer. This choice
of $\xi$ corresponds to a generalized beamsplitter
with a 50 \% transmittivity. We use the global maxima
and minima to define the visibility according to
(\ref{eq:quantum visibility definition}). It is
straightforward to show that from the obtained
expression follows that
\beq
V = 2 |\bra{A_+}
\hat{\rho}
\ket{A_-}| = 2 \rho_{12} .
\label{eq:visibility expressed in rho}
\eeq 
From this expression, and (\ref{eq:predictability})
follows that
\beq
P^2 + V^2 = w_+^2 - 2 w_+ w_- + w_-^2 + 4 \rho_{12}^2
\leq 1 ,
\label{eq:first complementarity}
\eeq
since $w_+ + w_- = 1$ and $0 \leq \rho_{12} \leq
\sqrt{w_+ w_-}$. This re-derivation of relation
(\ref{eq:Complementarity relation}) demonstrates
that every two-state quantum system obeys a
complementary relation even before any attempt has been
made to simultaneously measure both $\hat{A}$ and the
quantity complementary to $\hat{A}$. In Fig.
\ref{fig:Uncertainty product} the
generalized distinguishability and visibility are
plotted dash-dotted and long dashed, respectively, as
a function of $w_+$, parameterized by
$\sin^2\alpha = w_+$.

It is obvious that $P$ in some way corresponds to a
measurement of $\opa$. What is the operator
corresponding to the quantity represented by the
visibility? To answer this question let us construct
the two orthonormal states
\beq
\ket{B_+} \equiv (\ket{A_+} + e^{i \varrho}
\ket{A_-})/\sqrt{2}
\label{eq:b+ state definition}
\eeq
and
\beq
\ket{B_-}
\equiv (\ket{A_+} - e^{i \varrho} \ket{A_-})/\sqrt{2} .
\label{eq:b- state definition}
\eeq
These states, in turn, can be used to construct a
complementary Hermitian operator to
$\hat{A}$ which is
\beq
\hat{B} = B_+ \ket{B_+}\bra{B_+} + B_-
\ket{B_-}\bra{B_-},
\label{eq:b definition}
\eeq
where $B_+ \neq B_-$ are real numbers. We have denoted
this operator the complementary operator to
$\hat{A}$ since the eigenstates of $\opb$ are equally
weighted superpositions of the eigenstates of $\hat{A}$.
Hence, if we prepare a state so that the outcome of a
measurement corresponding to $\hat{A}$ can be predicted
with certainty, then nothing can be predicted about the
measurement outcome corresponding to $\hat{B}$,
and vice versa. Since the eigenstates of $\opb$ are
parameterized by $\varrho$ there exists a whole set of
complementary operators to $\opa$. We note that
irrespective of
$\varrho$, the states
$\ket{B_+}$ and
$\ket{B_-}$ are the Hadamard transformations of the
$\ket{A_+}$ and
$\ket{A_-}$ states. This is no coincidence since the
fact that $\hat{B}$ is the complementary operator
to $\hat{A}$ makes the two corresponding bases
optimal for quantum cryptography. This brings in light
the intimate link \cite{Fuchs} between the
complementarity relations, quantum information and
quantum cryptography.

It should be noted that the observable $\hat{A}$ plays
no particular role vis-a-vis the observable $\hat{B}$
in the expression (\ref{eq:first complementarity}).
Assume that the state $\hat{\rho}$ remains invariant,
but that $\opa$ and $\opb$ are interchanged so that the
maximum likelihood estimation and the
generalized visibility measurement pertains to the
outcome of a measurement of
$\opb$, and that new unitary transformations
are defined that have identically the same form as
(\ref{eq:ups}) and (\ref{eq:ubs}) if expressed in the
$\ket{B_+}$ and $\ket{B_-}$ basis. Then we find that
\beq
P_B = 2 \rho_{12} | \cos(\theta-\varrho) |
\label{eq:B predictability}
\eeq 
(where we have labeled this predictability with an
index $B$ not to confuse it with the predictability of
estimating $\hat{A}$) and that the new ``visibility''
is given by
\beq
V_B = \sqrt{w_+^2 + w_-^2 - 2 w_+ w_- + 4 \rho_{12}^2
\sin^2(\theta -\varrho)} .
\label{eq:B visibility}
\eeq 
Hence, although the likelihood of estimating $\opb$
correctly in general is different from the likelihood
of estimating $\opa$, relation (\ref{eq:first
complementarity}) still holds. We observe that it is
always possible to find an operator $\hat{B}$ for
which the predictability between the measurement
outcomes is zero. This represents the quantum erasure
measurement operator
\cite{Walther,Tan,Kwiat2,Bjork,Herzog,Scully,Kim}. However, the
sum
$P^2 + V^2$ remains invariant and depends only on the
state, not on the choice of complementary operators by
which one estimates and measures $P$ and $V$.

To explore the symmetry between the pairs $\opa$ and
$\opb$, we can see from
(\ref{eq:visibility expressed in rho}) and
(\ref{eq:B predictability}) that the
predictability
$P_B$ of a measurement of the {\em proper}
$\opb$ operator is 
\beq
P_B = 2 \rho_{12} = V
\label{eq:B and P relation}
\eeq
By the ``proper $\opb$ operator'' we mean the
complementary operator to $\hat{A}$ that, for the state
$\hat{\rho}$, optimizes the visibility (or
minimizes the variance of $\opb$, see below).
Therefore the operator is defined with
$\varrho=\theta$. Note that the word proper hence is
in reference to the measured state. In the same manner
we see from (\ref{eq:predictability}) and (\ref{eq:B
visibility}) that for the proper $\opb$ operator
\beq
V_B = |w_+ - w_-| = P .
\label{eq:B and V relation}
\eeq

\section{The Heisenberg-Robertson uncertainty relation}

The commutator between $\opa$ and $\opb$ follows from
the definition of the operators and can be expressed:
\beq
[\opa,\opb] = (A_+ - A_-)(B_+ - B_-)(\ket{A_+}\bra{A_-}
- \ket{A_-}\bra{A_+}) .
\label{eq:commutator}
\eeq
We see that the operators $\opa$ and $\opb$ are
non-commuting, non-canonical operators. Since the
operators are non-canonical they will be subject to a
generalized uncertainty inequality
\cite{Schrodinger,Robertson,Merzbacher,Puri,Trifonov}.
 The uncertainty inequality reads
\beq
\langle (\Delta \opa)^2 \rangle \langle(\Delta \opb)^2
\rangle \geq {1 \over 4} \left ( \langle \hat{C}
\rangle^2 + \langle \hat{F} \rangle^2 \right ) ,
\label{eq:uncertainty relation}
\eeq
where the $\hat{C}$ is directly proportional
to the commutator and is defined as
\beq
\hat{C} = -i [\opa,\opb] ,
\label{eq:C definition}
\eeq
and $\langle \hat{F} \rangle$ is the correlation between
the observables and is defined
\beq
\langle \hat{F} \rangle = \langle \opa \opb +
\opb \opa\rangle - 2 \langle \opa
\rangle \langle \opb\rangle .
\label{eq:F definition}
\eeq

The expectation value and the variance of the state
(\ref{eq:state definition}) can be computed to be
\beq
\langle \opa \rangle = A_+ w_+ + A_- w_- ,
\label{eq:expectaton value of a}
\eeq
and
\beq
\langle (\Delta \opa)^2 \rangle = (A_+ - A_-)^2 w_+ w_-
.
\label{eq:variance of a}
\eeq
Using (\ref{eq:predictability}) we can write
\beq
{\langle (\Delta \opa)^2 \rangle \over (A_+ - A_-)^2} = 
{1-P^2 \over 4}.
\label{eq:variance of a, alternative form}
\eeq
This equation shows the direct link between a
measurement of the operator $\opa$ and the
predictability $P$. If e.g. $P$ is unity, then the
variance of $\opa$ is zero.

To show that a similar relation holds between $\opb$
and $V$ we compute:
\begin{eqnarray}
\langle \opb \rangle & = & {B_+ \over 2} \, [ 1+ 2
\rho_{12} \cos(\theta-\varrho) ] \nonumber \\
& & + {B_- \over
2} \, [ 1- 2 \rho_{12} \cos(\theta-\varrho) ] ,
\label{eq:expectaton value of b}
\end{eqnarray}
and
\beq
\langle (\Delta \opb)^2 \rangle = (B_+ - B_-)^2 {1-4
\rho_{12}^2 \cos^2(\theta-\varrho)\over 4} .
\label{eq:variance of b}
\eeq
The variance of $\hat{B}$ is minimized for the
proper operator $\opb$. We note that for
the proper complementary operator, the normalized and
dimensionless variance is given by
\beq
{\langle (\Delta \opb)^2 \rangle \over (B_+ - B_-)^2} =
 {1-4 \rho_{12}^2\over 4} = {1-V^2 \over 4}.
\label{eq:normalized variance of b}
\eeq
The visibility is thus directly linked to the variance,
or uncertainty, of the proper complementary
operator to $\hat{A}$. As a final result we note that
the normalized uncertainty product can be written
\begin{eqnarray}
{\langle (\Delta \opa)^2 \rangle \langle (\Delta
\opb)^2 \rangle \over (A_+ - A_-)^2 (B_+ - B_-)^2} &
\geq & w_+ w_- {1-4 \rho_{12}^2 \over 4} = \nonumber \\
& & {(1-P^2)(1-V^2) \over 16} .
\label{eq:uncert prod, alternative form}
\end{eqnarray}
{\em Hence, there is a direct link between the
complementarity relation (\ref{eq:Complementarity
relation}) and the minimum uncertainty product.} This
is one of the main conclusions of this paper. The
result is true regardless if the state in question is
pure or mixed. However, for a pure state the right
hand side of (\ref{eq:uncert prod, alternative form})
can be further simplified to read 
\beq
{\langle (\Delta \opa)^2 \rangle \langle (\Delta
\opb)^2 \rangle \over (A_+ - A_-)^2 (B_+ - B_-)^2}
\geq {V^2 P^2 \over 16} . 
\label{eq:new uncertainty rel.}
\eeq
In Fig.
\ref{fig:Uncertainty product} the normalized minimum
uncertainty product (solid line) is plotted
versus the probability $w_+$. The dashed line
represents the maximum uncertainty product.

The Robertson intelligent states $\ket{\psi_{IS}}$ are
given by the eigenstate solution of the equation \cite{Trifonov}
\beq
(\opa + i \lambda \opb) \ket{\psi_{IS}} =
(\langle \opa \rangle + i \lambda
\langle \opb \rangle) \ket{\psi_{IS}} .
\label{eq:IS equation}
\eeq
The complex parameter $\lambda$ satisfies
$|\lambda|^2 = \langle (\Delta \opa)^2 \rangle/\langle (\Delta \opb)^2
\rangle$. We will here give the solutions for the cases where $\lambda$ is
either real or imaginary \cite{Puri}. For imaginary
$\lambda$ we find
\beq
\ket{\psi_{IS1}} = \sqrt{w_+} \ket{A_+} \pm e^{i
\varrho}
\sqrt{w_-} \ket{A_-} .
\label{eq. IS 1}
\eeq
As $|\lambda|$ increases from
zero towards infinity the two solutions evolve from
$w_+ = 1/2$ towards $w_+ \rightarrow 0$ and $w_+
\rightarrow 1$, respectively (from point A to points
B$_1$ and B$_2$ in Fig. \ref{fig:Uncertainty product}).
The minimum uncertainty states (which are the
intelligent states with the minimum uncertainty)
belong to this class of intelligent states with
$w_+ = 0$, 1/2, and 1, respectively. As expected these are the
eigenstates of
$\opa$ and $\opb$. The intelligent states
with $\lambda$ real are given by
\beq
\ket{\psi_{IS2}} = {1 \over \sqrt{2}}( \ket{A_+} \pm
e^{i (\varrho \pm \beta)}
\ket{A_-}) . 
\label{eq. IS 2a}
\eeq
When $|\lambda|$ goes from 0 towards 1,
$\beta$ goes from 0 to $\pi/2$ (and the uncertainty
product goes along the dotted line from point A to
point C in Fig.
\ref{fig:Uncertainty product}). When subsequently
$1 \leq |\lambda|
\rightarrow \infty$, the second class of intelligent
states becomes 
\beq
\ket{\psi_{IS2}} = \sqrt{w_+} \ket{A_+} \pm i e^{i
\varrho} \sqrt{w_-}
\ket{A_-} , 
\label{eq. IS 2b}
\eeq
where $w_+$ evolves from 1/2 towards 0 and from 1/2
towards 1, for the respective states. (The uncertainty
product goes along the dashed line from point C to
points B$_1$ and B$_2$.) The two sets of intelligent states considered
above are found to be the states with extreme uncertainty products.

\section{Complementary two-state operators}

In the preceding section we saw that there are
infinitely many complementary operators to
$\opa$. {\em However, for any choice of $\varrho$ there
are only two sets of three mutually complementary
operators.} If we explicitly write out the operator
$\hat{B}$ as a function of $\varrho$, v.i.z.
$\hat{B}(\varrho)$, then the two sets are
$\opa$, $\opb(\varrho)$, $\opb(\varrho + \pi/2)$ and
$\opa$, $\opb(\varrho)$, $\opb(\varrho - \pi/2)$. To
clarify the meaning of this statement we note that the
operator pair $\opa$ and $\opb(\varrho)$,  the
pair $\opa$ and
$\opb(\varrho+\pi/2)$ as well as the pair
$\opb(\varrho)$ and $\opb(\varrho+\pi/2)$ are all
complementary. This is the meaning of the term
``a set of mutually complementary operators''. 

To identify such an abstract
set of operators with one (of many) specific
observables we can e.g. assume that
$\opa$ corresponds to the spin
$z$ operator $\hat{\sigma}_z$ of a spin 1/2 particle,
with eigenvalues $\pm \hbar/2$. If we furthermore
assume that the eigenvalues of the operators
$\opb(\varrho)$ and $\opb(\varrho \pm \pi/2)$ are $B_1
= - B_2 = \hbar/2$, we readily recognize the other two
complementary operators as the spin
$\hat{\sigma}_u$ and
$\hat{\sigma}_v$ operators, where $u,v,z$ denotes a
right-handed (left-handed) orthogonal Euclidian vector
set for the choice $\varrho + \pi/2$ ($\varrho -
\pi/2$). With this choice the commutation relation
(\ref{eq:commutator}) reduces to the familiar spin
operator commutator
$[\hat{\sigma}_i,\hat{\sigma}_j] = i \hbar
\epsilon_{ijk} \hat{\sigma}_k$ with $i,j,k \rightarrow
u,v,z$, in this case. 

Note, however, that orthogonal
spin operators is only one realization of a mutually
complementary set. One can construct infinitely many
such triplets of operators for every two-state system.
In \cite{Bruss,Gisin} such a triplet set was
constructed starting from two orthogonally polarized
single photon states. The three pairs of eigenstates
were subsequently used as bases in a six-state quantum
cryptography protocol. {\em We think that a
generalization of this idea of mutually complementary
operators in Hilbert-spaces of higher dimension have
an obvious application for the construction of
eavesdropping-safe quantum cryptographic protocols.}

\begin{figure}
\leavevmode
\epsfxsize=8cm
\epsfbox{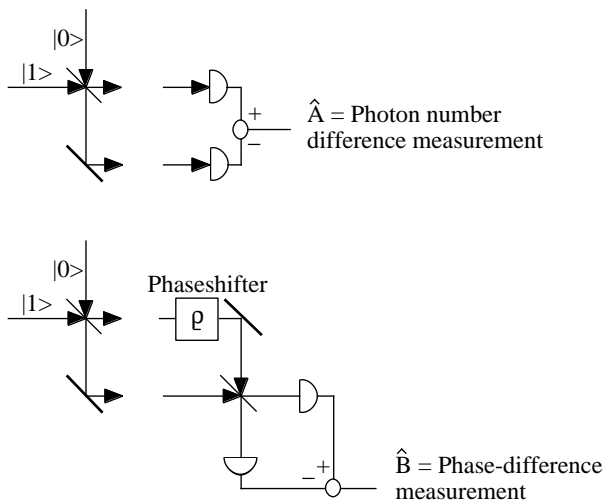}
\vspace{0.4cm}
\caption{An example of complementary operators. The
single photon two-mode state prepared at left can be
measured either by a number-difference operator (top)
or by a phase-difference measurement (bottom). By
adjusting the phase shift $\varrho$ any operator
$\opb(\varrho)$ can be implemented.}
\label{fig:Operator illustration}
\end{figure}

As a second example of complementary operator
pairs we consider a single particle in a well defined
spatio-temporal mode, impinging on a symmetric
Mach-Zehnder interferometer. The particle can take
either of the two paths. The corresponding
states can be written
$\ket{1}\otimes\ket{0}$ and $\ket{0} \otimes \ket{1}$.
Defining $\opa$ to be the two-mode particle
number-difference operator
$(\hat{n}_1 -\hat{n}_2)/2$ with the
eigenvectors above and with eigenvalues
$\pm 1/2$, we find that the operator $\opb$ corresponds
to the two-mode phase-difference operator (in the first
particle manifold) defined by Luis and S\'{a}nchez-Soto
\cite{Luis,Luis 2,Luis 3}:
\beq
\hat{\phi}_{12} = \theta |\phi_1\rangle\langle\phi_1| +
(\theta + \pi) |\phi_2\rangle\langle\phi_2|,
\label{eq:phase-difference states}
\eeq
where $\ket{\phi_1}=(\ket{1}\otimes\ket{0} +
e^{i \varrho} \ket{0} \otimes \ket{1})/\sqrt{2}$
and $\ket{\phi_2}=(\ket{1}\otimes\ket{0} - e^{i \varrho}
\ket{0} \otimes \ket{1})/\sqrt{2}$, and the
corresponding eigenvalues are $\theta$ and $\theta +
\pi$. Fig. \ref{fig:Operator illustration} provides a
schematic illustration of the implementation of the
state preparation (left) and the operator set (right).

{\em Note that the proper operators in this case,
where the eigenstates are discrete, do not correspond
to position and momentum operators}, as textbook
discussions of this specific interferometric duality
experiment often indicate. This has already been noted
and discussed by de Muynck \cite{Muynck}. In the same
vein Luis and S\'{a}nchez-Soto have used the
phase-difference operator to analyze the
mechanism which enforces complementarity, and
specifically loss of fringe visibility with increasing
distinguishability \cite{Luis 3}.

Identifying the operator $\opa$ with any Hermitian
operator (for which it makes sense to have a
restricted two-state Hilbert-space) it is
always possible to write down a complementarity
relation for $\opa$. Since the form
of the complementary operator $\opb$ to $\opa$ is
known, and it has a simple form, it is often possible
to identify this operator with some known observable.
Since complementarity follows directly from the
superposition principle, and therefore permeates all of
quantum mechanics, the term {\em welcher weg}
experiment which is intimately tied to complementarity
should perhaps be replaced by the term {\em welcher
zustand} experiment.

\section{Complementarity and uncertainty relations for
simultaneous measurements}

So far we have discussed the standard
Heisenberg-Robertson uncertainty relation which makes
a statement about the preparation of a state. E.g. the
variance of $\opa$ computed in (\ref{eq:variance of a})
is the variance associated with a
sharp measurement of $\opa$ on an ensemble of systems
all prepared in the state $\hat{\rho}$. The
measurement will either destroy the state or
collapse the state into an eigenstate of $\opa$ so the
sharp measurement will preclude any meaningful
subsequent measurement of $\opb$. However, it is also
interesting to see what the uncertainty product becomes
if one tries to {\em simultaneously} measure
$\opa$ {\em and} $\opb$. By ``simultaneous'' we
mean that we make (necessarily unsharp)
measurements of both $\opa$ and $\opb$ on each
individual system mode in the ensemble. In order to do
so we need to entangle the system with an auxiliary
meter mode, associated with an arbitrarily large
Hilbert-space
${\cal H}_m$, see Fig. \ref{fig:Simultaneous
measurement}. If the entanglement between the
system and meter modes is perfect (to be quantified
below), then a sharp measurement of the state of the
meter mode will collapse the system mode into an
eigenstate of e.g. $\opa$. This is the principle of a
quantum non-demolition measurement. However, in order
to subsequently be able to say something about $\opb$ of
the initial system state from the same copy of the
quantum state, the entanglement cannot be perfect, and
to  be able to say something about
$\opa$ it cannot be zero. Therefore both the
$\opa$ and the $\opb$ measurements need to be unsharp,
that is, associated with additional statistical
uncertainties than what follows from the preparation of
the state. This is well known for simultaneous
measurements \cite{Muynck,Arthurs,Arthurs
2,She,Stenholm,Appleby}.

\begin{figure}
\leavevmode
\epsfxsize=8cm
\epsfbox{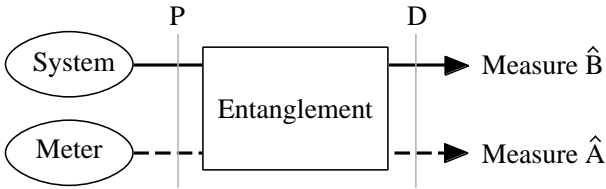}
\vspace{0.3cm}
\caption{A schematic representation of a simultaneous
measurement of $\opa$ and $\opb$ on a single system.
The planes P and D denote the spatial points
where it is appropriate to measure $P$ or $V$, and
$D$ or $V_e$, respectively. If an unsharp measurement
is performed $D$ {\em and} $V_e$ can be measured at
plane D.}
\label{fig:Simultaneous measurement}
\end{figure}

In the following we will only treat the case where the
composite system is pure as this will set the lower
limit to our ability to simultaneous measure or predict
the values of the observables
$\opa$ and $\opb$. We shall assume that the
entanglement is accomplished through some unitary
operation which does not change the probabilities $w_+$
and $w_-$. (The more general case is discussed in
\cite{Bjork}). Such an entanglement should perhaps be
called quantum non-demolition measurement
type entanglement (where the word measurement refers
to $\opa$), since a perfect entanglement of this type
will allow one to make a sharp QND measurement of
$\opa$. This is not the most general form of
entanglement, but it is the type of entanglement
needed for our purposes, so a sufficiently
general pure state of the entangled system can be
written
\begin{eqnarray}
\ket{\psi_e} & = & \sqrt{w_+} \ket{A_+} \otimes
\ket{M_+} + e^{i \theta} \sqrt{w_-} \ket{A_-} \otimes
\ket{M_-}
\nonumber \\ & \equiv & \sqrt{w_+} \ket{A_+} \otimes
\ket{M_+} \nonumber \\
& & + e^{i \theta} \sqrt{w_-} \ket{A_-} \otimes (c
\ket{M_+} +
\sqrt{1-c^2} \ket{M_\perp} ) ,
\label{eq:entangled system}
\end{eqnarray}
where $c$ is real and positive and defined by
$|\bra{M_-}M_+\rangle|=c$, and
$\bra{M_+}M_\perp\rangle=0$. We see that
since
$\opa$ has a binary measurement outcome, we need only
consider the reduced 2-dimensional meter mode
Hilbert-space ${\cal H}_r$ spanned by $\ket{M_+}$ and
$\ket{M_\perp}$. The density operator associated with
$\ket{\psi_e}$ will be denoted
$\hat{\rho}_e$. With the help of the parameter
$c$, which is a measure of the entanglement between the
modes, the distinguishability
$D$ based on an optimal measurement of the meter
mode, can be expressed
\beq
D \equiv ||\bra{A_+} \hat{\rho}_e \ket{A_+} - \bra{A_-}
\hat{\rho}_e
\ket{A_-}||_1 = \sqrt{1 - 4 c^2 w_+ w_-} .
\label{eq:distinguishability definition}
\eeq
(The notation $ ||\hat O||_1\equiv {\rm Tr}\{\sqrt{\hat
O^\dagger \hat O}\}$ denotes the trace-class norm of
the operator $\hat O$.) The distinguishability $D$ has
the same connection to the likelihood of guessing
correctly about the outcome of a sharp $\opa$
measurement, that is $D = 2L-1$. However, $D$ is
associated with a factual {\em measurement} and not
simply by an estimate based on {\em a priori}
knowledge. It may be noted that $P \leq D$ always holds
\cite{Englert,Bjork} provided the QND-type of
entanglement is used.

The visibility $V_e$ of the entangled system is given
by
\beq
V_e = 2 |\bra{A_-} {\rm Tr}_r \{ \hat{\rho}_e \}\ket{A_+}| =
2 c \sqrt{w_+ w_-} ,
\label{eq:visibility definition}
\eeq
where the partial trace is taken over ${\cal H}_r$.
Since the state $\ket{\psi_e}$ is pure, the relation
\beq
D^2 + V_e^2 =1
\label{eq:D complementarity}
\eeq 
holds \cite{Englert}.

A measurement of the meter mode will allow us to
deduce more or less about the initial state of the
system (\ref{eq:state definition}). If e.g. $c=0$,
then the wavefunction (\ref{eq:entangled system})
represents a maximally entangled state and a
measurement of the meter mode, using
$\ket{M_+}$ and
$\ket{M_\perp}$ as projection bases, will yield a
perfect correlation with a subsequent $\opa$ measurement
of the system mode. However, as noted above, such a
measurement of the meter mode will preclude any
meaningful information to be reaped about the
complementary observable $\opb$. Therefore, we will in
the following discuss how to perform an optimal
simultaneous measurement of
$\opa$ and $\opb$. With simultaneous we understand a
measurement where we try to obtain some information of
both $\opa$ and $\opb$ from a single copy of a
pure state (\ref{eq:state definition}) which after an
entangling interaction is transformed to
(\ref{eq:entangled system}).

To find the minimum uncertainty product of a
simultaneous measurement of $\opa$ and $\opb$ we
have assumed that the $\opa$ information will be
obtained by making a sharp measurement of the meter
mode, and the
$\opb$ information will be obtained by making a
subsequent sharp measurement on the system mode. (We
note that one could also have chosen to do the
opposite. However, if we do so, the system and meter
modes should be entangled in a different manner than
what has been assumed above. Still, if we choose the
inverse measurement procedure and do it optimally, the
final result, quantified by a simultaneous uncertainty
product, remains identical to the result below.) Let us
start with the
$\opb$ measurement. The two pertinent projectors are
$|B_+\rangle \langle B_+|$ and $|B_-\rangle \langle 
B_-|$. The associated probabilities are
\beq
P_{B_\pm} = \bra{B_\pm} {\rm Tr}_r \{ \hat{\rho}_e \}
\ket{B_\pm} = {1
\over 2} \pm c \sqrt{w_+ w_-} \cos(\theta-\varrho) .
\label{eq:PB+}
\eeq
If the two outcomes are associated with the
values
$B_+'$ and
$B_-'$, then the mean of the measurement, or rather
estimation, of
$\opb$ (which we will denote $\opb'$) will yield
\beq
\langle \opb' \rangle = {B_+' + B_-' \over 2} + (B_+' -
B_-') c
\sqrt{w_+ w_-}
\cos(\theta-\varrho) .
\label{eq:Mean B estimation}
\eeq
We note that, without loss of generality, we can
assume a gauge such that the true eigenvalues $B_+$
and $B_+$ fulfill $B_+ = - B_- = B$, and hence $B_+' = -
B_-' = B'$. To make the estimated mean (\ref{eq:Mean B
estimation}) equal the true mean (\ref{eq:expectaton
value of b}) we set $B_\pm' = B_\pm/c$. We point out
that the choice is independent of the initial state of
the system mode (which is characterized by $w_+$ and
$\theta$), but depends on the degree of entanglement
$c$. The assumption of a ``true mean'' meter is
essential to what follows, and is also made by Arthurs,
Kelly and Goodman in their seminal papers on
simultaneous measurements \cite{Arthurs,Arthurs 2}.
From the assumption follows that the variance of the
estimate of $\opb$ is
\begin{eqnarray}
\langle (\Delta \opb')^2 \rangle & = & {(B_+'-B_-')^2
\over 4} \, [ 1 - 4 c^2 w_+ w_-
\cos^2 (\theta-\varrho) ] \nonumber \\
& = & B^2 \left( {1 \over c^2} - 4
w_+ w_-
\cos^2 (\theta-\varrho) \right) .
\label{eq:B estimation variance}
\end{eqnarray}
It is obvious from this expression that in order to
minimize the variance of the estimate of $\opb$, one
should choose $\theta=\varrho$. We also see that when
$c =1$, that is, when the system and meter modes are
unentangled, the result reduces identically to
(\ref{eq:variance of b}). On the other hand, when
$c\rightarrow 0$, the variance diverges in spite
of the fact that
$\opb$ has a finite number of finite eigenvalues.
This is a consequence of our requirement that the
mean of the estimate should equal the true mean of the
state.

If the (proper) choice $\theta=\varrho$ is made, then
we can express the normalized variance in the
visibility (\ref{eq:visibility definition}):
\beq
{\langle (\Delta \opb')^2 \rangle \over (B_+'-B_-')^2}
= {1-V_e^2 \over 4}
\label{eq:reformulation}
\eeq
It is to be expected that such a relation holds,
because the visibility, as shown above, corresponds to
a sharp measurement of the uncertainty of operator
$\opb$.

In order to best estimate the outcome of a 
measurement of $\opa$ of the initial system state from a
measurement on the meter mode, we need to find the
optimal projectors. Since the entangled state
(\ref{eq:entangled system}) can be expressed in a 2
$\times$ 2 Hilbert space, we need only construct two
meter mode projectors. The most general forms for the
projector-states are
\beq
\ket{M_1} = \cos(\gamma) \ket{M_+} + e^{i \kappa}
\sin(\gamma) \ket{M_\perp}
\label{eq:Def M1}
\eeq
and
\beq
\ket{M_2} = -\sin(\gamma) e^{i \kappa} \ket{M_+} +
\cos(\gamma) \ket{M_\perp} .
\label{eq:Def M2}
\eeq
However, it is immediately obvious that in order to
best estimate $\opa$ the choice $\kappa=0$ (or
$\kappa=\pi$ which is equivalent to $\ket{M_1}
\leftrightarrow \ket{M_2}$) should be made since
$\ket{M_\perp}$ is defined such that $c$ is real.
Furthermore, again we shall assume that a gauge is
chosen so that $A_+ = -A_- = A$, and therefore the
measurement values associated with the two outcomes
$A_+'$ and $A_-'$ fulfill $A_+' = -A_-' = A'$. With this
choice, the expectation value of the estimate of
$\opa$ (which we call $\opa'$) becomes
\begin{eqnarray}
\langle \opa' \rangle & = & A' \{ [\cos^2(\gamma)
- \sin^2(\gamma)] [ w_+ - w_-(1-2c^2)]  \nonumber
\\ & &  + 4 w_- c
\sqrt{1-c^2}
\cos(\gamma)
\sin (\gamma) \} 
\label{eq:Mean A estimation}
\end{eqnarray}
We see that in order to make the estimated mean
correct, c.f. (\ref{eq:expectaton value of a}),
and independent of the initial system state, the
following two conditions must be met:
\beq
{\cos^2(\gamma) - \sin^2(\gamma) \over
2 \cos(\gamma) \sin(\gamma) } = -{\sqrt{1-c^2} \over
c}
\label{eq:Condition 1}
\eeq
and
\beq
A' = {A \over \cos^2(\gamma) - \sin^2(\gamma)} = {A
\over \sqrt{1-c^2}},
\label{eq:Condition 2}
\eeq
where the second equality in (\ref{eq:Condition 2})
follows from (\ref{eq:Condition 1}). Note that both
conditions are state independent. That is, $\langle
\opa' \rangle = \langle \opa \rangle$ irrespective of
$w_+$ and $\theta$ if the two conditions are met. The
variance of the estimate of
$\opa$ can then be computed to be
\beq
\langle (\Delta \opa')^2 \rangle = A^2 \left ( {c^2
\over 1-c^2} + 4 w_+ w_- \right ).
\label{eq:A estimation variance}
\eeq
We see that, as expected, the estimated variance equals
the true variance for $c=0$, that is, when the system
and meter modes are maximally entangled. The estimated
variance diverges when $c \rightarrow 1$. This, too,
is a consequence of requiring a correct estimated mean.

The normalized and dimensionless simultaneous
uncertainty product is hence
\beqa
\lefteqn{{\langle (\Delta \opa')^2 \rangle \langle (\Delta
\opb')^2 \rangle \over 16 A^2 B^2}} & & \nonumber \\
& & = \left ( {c^2 \over
4(1-c^2)} + w_+ w_- \right )\left ( {1 \over 4 c^2} - 
w_+ w_-
\right ).
\label{eq:Uncertainty product}
\eeqa
As expected, the uncertainty product is larger
than (\ref{eq:uncert prod, alternative form}) and
depends both on the initial state and on the degree of
entanglement between the system mode and the meter
mode. The entanglement parameter $c$ can be used to
shift the measurement uncertainty, to some extent, from
one of the operators to the complementary one. 

\begin{figure}
\leavevmode
\epsfxsize=8cm
\epsfbox{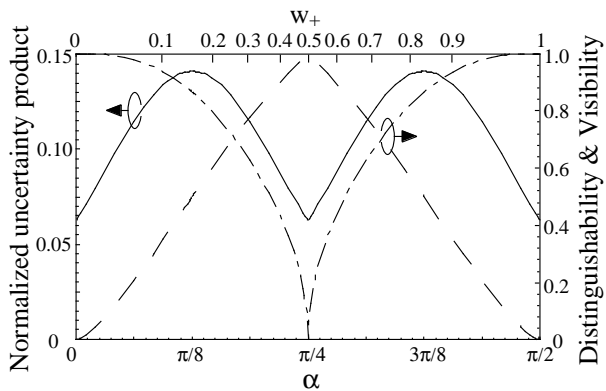}
\vspace{0.3cm}
\caption{The normalized dimensionless
minimum uncertainty product for a
simultaneous measurement of the operators $\opa$ and
$\opb$, solid line. The dot-dashed line represents the
distinguishability and the long dashed line the
visibility of the entangled state, provided that the
optimum entanglement parameter $c$ is chosen for each
value of $w_+$.}
\label{fig:Simultaneous uncertainty product}
\end{figure}

If, for each choice of
$w_+$, we optimize the entanglement to minimize the
uncertainty product, and this is the principle used
in \cite{Arthurs,Arthurs 2} to derive the
minimum uncertainty product for a simultaneous
measurement of two {\em canonical} non-commuting
operators, then the ensuing normalized minimum
uncertainty product is given by:
\end{multicols}
\vspace{-0.5cm}
\noindent\rule{0.5\textwidth}{0.4pt}\rule{0.4pt}{\baselineskip}
\widetext
\beq
{\langle (\Delta \opa)^2 \rangle \langle (\Delta
\opb)^2 \rangle \over 16 A^2 B^2} = {-16 w_+^2 
w_-^2 (1-4w_+ w_-)^2+ [ 1 - 12 w_+ w_-(1-4w_+ w_-) ]
\sqrt{ w_+ w_- (1 - 4  w_+ w_-)} \over 16 [ -4 w_+ 
w_- (1-4w_+ w_-) + \sqrt{w_+ w_- (1 - 4 w_+ 
w_-)} ]}
\label{eq:Messy expression}
\eeq
\begin{multicols}{2}
\noindent
and this expression holds for
\beq
c = \sqrt{{-4 w_+  w_- + 2\sqrt{w_+  w_-(1-4w_+  w_-)}}
\over 1 - 8 w_+ w_- } .
\label{eq:optimal entanglement}
\eeq
This is the other major result of this paper,
where the physical implications of (\ref{eq:Messy
expression}) rather than the (rather messy) form
should be retained. Remember that this result is
contingent on {\em a priori} information about the
preparation of the state (i.e. to be able to make a
minimum uncertainty measurement of the state,
$w_+$ and
$\theta$ must be known). The result is plotted in Fig.
\ref{fig:Simultaneous uncertainty product},
solid line. This is the minimum uncertainty of a
simultaneous measurement of the two complementary and
non-canonical operators $\opa$ and $\opb$. The result
should be compared with the standard uncertainty
product (\ref{eq:uncert prod, alternative form}) of
a state in two-dimensional Hilbert-space.

The corresponding distinguishability and visibility
are plotted as dash-dotted and long dashed lines,
respectively, in Fig. \ref{fig:Simultaneous uncertainty
product}. We have assumed that the entanglement
parameter $c$ for each
$w_+$ is chosen according to (\ref{eq:optimal
entanglement}). By necessity the distinguishability
is higher than the predictability for the entangled
state while the visibility is correspondingly lower
than for the initial state (c.f. Fig.
\ref{fig:Uncertainty product}). We
also observe that due to the ``noise term'' in
(\ref{eq:A estimation variance}) there is no simple
connection between the uncertainties of $\opa'$ and
the distinguishability (\ref{eq:distinguishability
definition}). Therefore one cannot find any general
relation between (\ref{eq:D complementarity}) and the
non-simultaneous uncertainty relation (\ref{eq:uncert
prod, alternative form}). This was noted already by
Englert who asserted that ``\ldots the duality
relation (\ref{eq:D complementarity}) is logically
independent of the uncertainty relation
\ldots .'' The reason is that whereas $\langle (\Delta
\opa)^2 \rangle$, $\langle (\Delta \opb)^2 \rangle$,
$P$, $V$ and $V_e$ all correspond to the uncertainties
of factual measurements, $D$ does not. Instead it
characterizes a ML estimate, which cannot be directly
related to a factual measurement of the
corresponding operator.

It is interesting to make a connection between
(\ref{eq:Messy expression}) and (\ref{eq:uncert prod,
alternative form}) via (\ref{eq:optimal
entanglement}). We have already noted that the two
uncertainty relations are different since they
represent physically different measurements on
physically different states. However, we have striven,
as far as possible, to try to design a ``fair''
procedure to simultaneous measure
$\opa$ and $\opb$ of the state (\ref{eq:state
definition}) through a measurement of the
state (\ref{eq:entangled system}). Specifically we
have required that the expectation values of
the corresponding measurements coincide. Using the fact
that the optimum entanglement given by (\ref{eq:optimal
entanglement}) can be expressed
\beq
c = \sqrt{{V(P-V) \over P^2-V^2}} ,
\label{eq:new optimal c}
\eeq
some somewhat tedious algebra will show that the
normalized simultaneous uncertainty product can
be expressed
\beq
{\langle (\Delta \opa')^2 \rangle \langle (\Delta
\opb')^2 \rangle \over 16 A^2 B^2} = {(1-VP)^2 \over
16}.
\label{eq:Uncertainty product 2}
\eeq
This rather remarkably simple result should be
compared to (\ref{eq:uncert prod, alternative
form}) and (\ref{eq:new uncertainty rel.}). {\em Hence,
we see that our requirement of correct expectation
values enforces a simultaneous uncertainty
product which is uniquely dictated by the preparation
of the state.} This is not wholly surprising as e.g.
the eigenstates of
$\opa$ and
$\opb$ could reasonably be expected to have the
smallest simultaneous uncertainty product.

Appleby \cite{Appleby} has argued that the
simultaneous uncertainty relation is less general than
the Heisenberg-Robertson uncertainty relation. We
agree, and this is demonstrated by our analysis. While
the Heisenberg-Robertson uncertainty relation is based
on sharp, non-simultaneous measurements of the two
complementary operators, and therefore operationally
well defined, the simultaneous uncertainty relation is
based on unsharp measurements. How should these unsharp
measurements be performed? More specifically, how
should e.g. the meter-state projection basis
(\ref{eq:Def M1}) and (\ref{eq:Def M2}) be chosen, and
how should the outcomes of the meter-system
measurement be interpreted? We have, following Arthurs
and Kelly, required that the expectation value of
the unsharp measurement equals the true mean. This
requirement enters our analysis through
(\ref{eq:Condition 1}) and (\ref{eq:Condition 2}). In
Appleby's terminology this defines a retrodictively
unbiased measurement. This is, however, not the only
reasonable choice. We can instead choose the
meter-state projection basis to optimize the
distinguishability for every value of the entanglement
parameter $c$. With this choice one cannot make a
state-independent retrodictively unbiased measurement.
None-the-less, the choice is reasonable but will
result in a different minimum uncertainty relation
than the one we derived. In our eyes, if the meter
projection basis is chosen to optimize the
distinguishability, it is more natural to make the
quantitative statement of complementarity in terms of
equation (\ref{eq:D complementarity}). The conclusion
is that any expression that makes a statement of a
simultaneous measurement of complementary observables
irrevocably must involve properties of the meter, in
addition to the properties of the state. 

It is noteworthy that the simultaneous uncertainty
relation we derived differs from that derived by
Arthurs, Kelly, and Goodman \cite{Arthurs,Arthurs 2}.
They, and subsequent workers, implicitly or
explicitly assumed that the pertinent operators were
canonical, and in this case the uncertainty product
for a simultaneous measurement is simply four times
larger than the standard uncertainty product. In our
case the situation is more complex. Due to the fact
that $\opa$ and $\opb$ are non-canonical the
uncertainty product of a simultaneous measurement is
not simply scaled by a constant factor. Specifically,
the simultaneous uncertainty product of the
eigenstates to
$\opa$ and $\opb$ is non-zero due to the ``correct mean
assumption''. We believe that this is a general result
for any non-canonical observables.

\section{Conclusions}

In physics textbooks quantum complementarity is often
exemplified in terms of one ``particle'' passing
through a double-slit. The complementary observables
are usually taken to be the canonical position and
momentum operators, without much justification. We
have shown how complementarity is a natural
consequence of the superposition principle, and
explored the connection between complementarity and
the uncertainty relations. We have shown that for any
two-state system one can always formulate a
generalized complementarity relation, and that this
relation typically cannot be interpreted in terms of
position and momentum operators. We have also indicated
that for a system with a discrete number of
non-degenerate eigenstates, the corresponding
operators are not canonical. Never-the-less, for a
two-state system simple and rather intuitive relations
hold between expressions of complementarity and
uncertainty. 

We have also shown that if a simultaneous
measurement of complementary operators is made on a
two-state system, a different uncertainty
relation arises than that derived by Arthurs
and Kelly. Finally, we have indicated some natural
connections between complementarity and quantum
information. This is to be expected since a two-state
system is a natural physical manifestation of a qubit,
irrespective of its particular physical implementation
(spin, excitation, charge, etc.).

\acknowledgments

This work was supported by grants from
the Swedish Technical Science Research Council, STINT,
the Royal Swedish Academy of Sciences and by INTAS
through Grant 167/96.

\end{multicols}

\end{document}